\documentclass[preprint,double-spaced,noshowpacs,preprintnumbers,amsmath,amssymb]{revtex4}

\usepackage{graphicx}% Include figure files
\usepackage{dcolumn}% Align table columns on decimal point
\usepackage{bm,slashed}% bold math
\usepackage{amssymb}
\begin{document}

\preprint{}

\title{The Dirac equation in a Yang-Mills field as an equation for just one real function}% Force line breaks with \\

\author{Andrey Akhmeteli}
 \email{akhmeteli@ltasolid.com}
 \homepage{http://www.akhmeteli.org}
\affiliation{%
LTASolid Inc.\\
10616 Meadowglen Ln 2708\\
Houston, TX 77042, USA}

%\affiliation{

\date{\today}% It is always \today, today,

\begin{abstract}
 Previously (A. Akhmeteli, J. Math. Phys., v. 52, p. 082303 (2011)), the Dirac equation in an arbitrary electromagnetic field was shown to be generally equivalent to a fourth-order equation for just one component of the four-component Dirac spinor function, and the remaining component can be made real by a gauge transformation. This work extends the result to the case of the Dirac equation in the Yang-Mills field.
\end{abstract}

\maketitle

\section{\label{sec:level1a}Introduction}

Schr{\"o}dinger ~\cite{Schroed} noticed that a charged scalar field can be made real by a gauge transformation. It was shown recently (~\cite{Akhmeteli-JMP,Akhm2015,Akhmspr}; see also ~\cite{Bagrov2014}, pp. 24-25,~\cite{Bagro}) that, in a general case, three out of four complex components of the Dirac spinor can be algebraically eliminated from the Dirac equation in an arbitrary electromagnetic field, and the remaining component can be made real by a gauge transformation, at least locally. However, the Dirac equation is also a key part of the Standard Model, so it is interesting to see if similar results can be derived for the Dirac equation in a non-Abelian gauge field. We consider the case of the Dirac equation in SU($n$) Yang-Mills field, following the more general approach of ~\cite{Akhm2015,Akhmspr}. All fields are classical (non-second-quantized) in this work. The fermion field has $4n$ components. It turns out that $3n$ components can be algebraically eliminated in a general case, and then, choosing a unitary gauge ~\cite{Itzykson}, all the remaining components but one can be made zero, and the last component can be made real.

\maketitle

\section{\label{sec:level1b}Algebraic elimination of spinor components from the Dirac equation in the Yang-Mills field}

Let us start with the Dirac equation in a Yang-Mills field (~\cite{Schwartz}, p.493)in the following form:
\begin{equation}\label{eq:pr25qcd}
(i\slashed{\partial}-\slashed{A})\psi=\psi,
\end{equation}
where, e.g., $\slashed{A}=A_\mu\gamma^\mu$ (the Feynman slash notation), $A_\mu=A_\mu^a T^a$, $T^a$ are generators of the SU($n$) group. For the sake of simplicity, a system of units $\hbar=c=m=1$ is used, and the charge is included in $A_\mu$. Components $\psi_{i\alpha}$ of the field $\psi$ have an SU($n$) fundamental representation index (``color" index) $i$ and a spinor index $\alpha$. The metric tensor used to raise and lower indices is ~\cite{Itzykson}
\begin{equation}\label{eq:d1ps}
\nonumber
g_{\mu\nu}=g^{\mu\nu}=\left( \begin{array}{cccc}
1 & 0 & 0 & 0\\
0 & -1 & 0 & 0\\
0 & 0 & -1 & 0\\
0 & 0 & 0 & -1 \end{array} \right), g_\mu^\nu=\delta_\mu^\nu.
\end{equation}
Multiplying both sides of equation (\ref{eq:pr25qcd}) by $(i\slashed{\partial}-\slashed{A})$ from the left and using notation
\begin{equation}\label{eq:li2}
\sigma^{\mu\nu}=\frac{i}{2}[\gamma^\mu,\gamma^\nu],
\end{equation}
\begin{eqnarray}\label{eq:d8n1}
F_{\mu\nu}=i[i\partial_\mu-A_\mu,i\partial_\nu-A_\nu]=A_{\nu,\mu}-A_{\mu,\nu}+i[A_\mu,A_\nu],
\end{eqnarray}
we obtain (cf. ~\cite{Schwartz}, p.173):
\begin{eqnarray}\label{eq:li1qcd}
\nonumber
\psi=(i\gamma^\nu\partial_\nu-A_\nu\gamma^\nu)(i\gamma^\mu\partial_\mu-A_\mu\gamma^\mu)\psi=
\\
\nonumber
(i\partial_\nu-A_\nu)(i\partial_\mu-A_\mu)\gamma^\nu\gamma^\mu\psi=
(i\partial_\nu-A_\nu)(i\partial_\mu-A_\mu)\frac{1}{2}(\{\gamma^\nu,\gamma^\mu\}+
[\gamma^\nu,\gamma^\mu])\psi=
\\
\nonumber
((i\partial_\mu-A_\mu)(i\partial^\mu-A^\mu)+\frac{1}{4}
(\{i\partial_\nu-A_\nu,i\partial_\mu-A_\mu\}+
[i\partial_\nu-A_\nu,i\partial_\mu-A_\mu])[\gamma^\nu,\gamma^\mu])\psi=
\\
\nonumber
((i\partial_\mu-A_\mu)(i\partial^\mu-A^\mu)+\frac{1}{4}
[i\partial_\nu-A_\nu,i\partial_\mu-A_\mu][\gamma^\nu,\gamma^\mu])\psi=
\\
(-\partial^\mu\partial_\mu-2 i A^\mu\partial_\mu-i A^\mu_{,\mu}+A^\mu A_\mu-\frac{1}{2}F_{\nu\mu}\sigma^{\nu\mu})\psi.
\end{eqnarray}
We used the fact that the contraction of a symmetric and an antisymmetric tensors vanishes.

Equation (\ref{eq:li1qcd}) is similar to an equation derived for the Dirac equation in electromagnetic field and used by Feynman and Gell-Mann ~\cite{Feygel} to eliminate two out of four components of the Dirac spinor function (see also an earlier article ~\cite{Laporte}). We obtain:
\begin{equation}\label{eq:li3}
(\Box'+F)\psi=0,
\end{equation}
where the modified d'Alembertian $\Box'$ is defined as follows:
\begin{eqnarray}\label{eq:d8n2}
\Box'=\partial^\mu\partial_\mu+2 i A^\mu\partial_\mu+i A^\mu_{,\mu}-A^\mu A_\mu+1=-(i\partial_\mu-A_\mu)(i\partial^\mu-A^\mu)+1,
\end{eqnarray}
and
\begin{equation}\label{eq:li4}
F=\frac{1}{2}F_{\nu\mu}\sigma^{\nu\mu}.
\end{equation}
Let us note that $\Box'$ and $F$ are manifestly relativistically covariant.

We assume that the set of $\gamma$-matrices satisfies the standard hermiticity conditions ~\cite{Itzykson}:
\begin{equation}\label{eq:li5a}
\gamma^{\mu\dag}=\gamma^0\gamma^\mu\gamma^0, \gamma^{5\dag}=\gamma^5.
\end{equation}
Then a charge conjugation  matrix $C$ can be chosen in such a way (~\cite{Bogol},~\cite{Schweber}) that
\begin{equation}\label{eq:li5}
C\gamma^\mu C^{-1}=-\gamma^{\mu T}, C\gamma^5 C^{-1}=\gamma^{5 T}, C\sigma^{\mu\nu}C^{-1}=-\sigma^{\mu\nu T},
\end{equation}
\begin{equation}\label{eq:li6}
C^T=C^\dag=-C, CC^\dag=C^\dag C=I, C^2=-I,
\end{equation}
where the superscript $T$ denotes transposition, and $I$ is the unit matrix.

We will consider ``spinor components" of the field $\psi$ having the form $\bar{\xi}\psi$, where $\xi$ is a Dirac spinor. Such ``spinor components" have ``color" components $\xi^*_\alpha\gamma^0_{\alpha\beta}\psi_{i\beta}$. Let us further assume that $\xi$ is a constant spinor  (so it does not depend on the spacetime coordinates $x=(x^0,x^1,x^2,x^3)$, and $\partial_\mu\xi\equiv 0$) and multiply  both sides of equation (\ref{eq:li3}) by $\bar{\xi}$ from the left:
\begin{equation}\label{eq:li7}
\Box'(\bar{\xi}\psi)+\bar{\xi}F\psi=0.
\end{equation}
To derive an equation for only one ``spinor component" $\bar{\xi}\psi$, we need to express $\bar{\xi}F\psi$ via $\bar{\xi}\psi$, but the author cannot do this for an arbitrary spinor $\xi$ or prove that this cannot be done. Therefore, to simplify this task, we demand that $\xi$ is an eigenvector of $\gamma^5$, in other words, $\xi$ is either right-handed or left-handed (the derivation can also be performed for such constant spinors multiplied by a function of spacetime coordinates). This condition is Lorentz-invariant. Indeed, Dirac spinors $\chi$ transform under a Lorentz transformation as follows:
\begin{equation}\label{eq:li7a}
\chi'=\Lambda \chi,
\end{equation}
where matrix $\Lambda$ is non-singular and commutes with $\gamma^5$ if the Lorentz transformation is proper and anticommutes otherwise ~\cite{Bogo}. Therefore, if $\xi$ is an eigenvector of $\gamma^5$, then $\xi'$ is also an eigenvector of $\gamma^5$, although not necessarily with the same eigenvalue.

Eigenvalues of $\gamma^5$ equal either $+1$ or $-1$, so $\gamma^5\xi=\pm\xi$. The linear subspace of eigenvectors of $\gamma^5$ with the same eigenvalue as $\xi$ is two-dimensional, so we can choose another constant spinor $\eta$ that is an eigenvector of $\gamma^5$  with the same eigenvalue as $\xi$ in such a way that $\xi$ and $\eta$ are linearly independent. This choice is Lorentz-covariant, as matrix $\Lambda$ in equation (\ref{eq:li7a}) is non-singular.

Obviously, we can derive an equation similar to (\ref{eq:li7}) for $\eta$:
\begin{equation}\label{eq:li7f}
\Box'(\bar{\eta}\psi)+\bar{\eta}F\psi=0.
\end{equation}

We can express $F_{\mu\nu}$ as $F_{\mu\nu}^a T^a$, where $T^a$ are the generators of the SU($n$) group, then $F=F^a T^a$, where $F^a=\frac{1}{2}F_{\nu\mu}^a\sigma^{\nu\mu}$. If $\gamma^5\xi=\pm\xi$, then $\bar{\xi}=\xi^\dag\gamma^0$ is a left eigenvector of $\gamma^5$ with an eigenvalue $\mp 1$, as
\begin{equation}\label{eq:li7b}
\bar{\xi}\gamma^5=\xi^\dag\gamma^0\gamma^5=-\xi^\dag\gamma^5\gamma^0=-(\gamma^5\xi)^\dag\gamma^0=\mp\bar{\xi}.
\end{equation}
The same is true for spinors $\bar{\eta}=\eta^\dag\gamma^0$ (the proof is identical to that in (\ref{eq:li7b})), $\bar{\xi}F^a$, and $\bar{\eta}F^a$, as $\gamma^5$ commutes with $\sigma^{\mu\nu}$ ~\cite{Itzykson}.
As the subspace of left eigenvectors of $\gamma^5$ with an eigenvalue $\mp 1$ is two-dimensional and includes spinors $\bar{\xi}F^a$, $\bar{\eta}F^a$, $\bar{\xi}$, and $\bar{\eta}$, where the two latter spinors are linearly independent (otherwise spinors $\xi$ and $\eta$ would not be linearly independent), there exist such $a^a=a^a(x)$, $b^a=b^a(x)$, $a'^a=a'^a(x)$, $b'^a=b'^a(x)$ that
\begin{equation}\label{eq:lin12qcd}
\bar{\xi}F^a=a^a\bar{\xi}+b^a\bar{\eta},
\end{equation}
\begin{equation}\label{eq:lin13qcd}
\bar{\eta}F^a=a'^a\bar{\xi}+b'^a\bar{\eta}.
\end{equation}

For each spinor $\chi$ the charge conjugated spinor
\begin{equation}\label{eq:lin14}
\chi^c=C\bar{\chi}^T
\end{equation}
can be defined, and it has the same transformation properties under Lorentz transformations as $\chi$ ~\cite{Schweber}. We have
\begin{equation}\label{eq:lin15}
\bar{\chi}\chi^c=\bar{\chi}C\bar{\chi}^T=(\bar{\chi})_\alpha C_{\alpha\beta}(\bar{\chi})_\beta=0,
\end{equation}
as $(\bar{\chi})_\alpha(\bar{\chi})_\beta$ and $ C_{\alpha\beta}$ are respectively symmetric and antisymmetric (see equation (\ref{eq:li6})) with respect to transposition of $\alpha$ and $\beta$.

Let us multiply equations (\ref{eq:lin12qcd}),(\ref{eq:lin13qcd}) by $\xi^c$ and $\eta^c$ from the right:
\begin{eqnarray}\label{eq:lin16qcd}
\nonumber
\bar{\xi}F^a\xi^c=a^a(\bar{\xi}\xi^c)+b^a(\bar{\eta}\xi^c)=b^a(\bar{\eta}\xi^c),
\\
\nonumber
\bar{\xi}F^a\eta^c=a^a(\bar{\xi}\eta^c)+b^a(\bar{\eta}\eta^c)=a^a(\bar{\xi}\eta^c),
\\
\nonumber
\bar{\eta}F^a\xi^c=a'^a(\bar{\xi}\xi^c)+b'^a(\bar{\eta}\xi^c)=b'^a(\bar{\eta}\xi^c),
\\
\nonumber
\bar{\eta}F^a\eta^c=a'^a(\bar{\xi}\eta^c)+b'^a(\bar{\eta}\eta^c)=a'^a(\bar{\xi}\eta^c),
\end{eqnarray}
so
\begin{eqnarray}\label{eq:lin17qcd}
a^a=\frac{\bar{\xi}F^a\eta^c}{\bar{\xi}\eta^c},
b^a=\frac{\bar{\xi}F^a\xi^c}{\bar{\eta}\xi^c},
a'^a=\frac{\bar{\eta}F^a\eta^c}{\bar{\xi}\eta^c},
b'^a=\frac{\bar{\eta}F^a\xi^c}{\bar{\eta}\xi^c}.
\end{eqnarray}
Let us note that
\begin{equation}\label{eq:lin18}
\bar{\xi}\eta^c=\bar{\xi}C\bar{\eta}^T=(\bar{\xi})_\alpha C_{\alpha\beta}(\bar{\eta})_\beta=-(\bar{\eta})_\beta C_{\beta\alpha}(\bar{\xi})_\alpha=-\bar{\eta}\xi^c
\end{equation}
and
\begin{equation}\label{eq:lin19qcd}
\bar{\xi}F^a\eta^c=\bar{\xi}F^a C\bar{\eta}^T=(\bar{\xi}F^a C\bar{\eta}^T)^T=\bar{\eta}C^T F^{aT}\bar{\xi}^T=\bar{\eta}F^a C\bar{\xi}^T=\bar{\eta}F^a\xi^c,
\end{equation}
as
\begin{equation}\label{eq:lin20}
\sigma_{\mu\nu}C=-C\sigma_{\mu\nu}^T
\end{equation}
(see equations (\ref{eq:li5}),(\ref{eq:li6})). Therefore,
\begin{equation}\label{eq:lin20aqcd}
b'^a=-a^a.
\end{equation}

Equations (\ref{eq:li7}), (\ref{eq:li7f}),(\ref{eq:lin12qcd}), (\ref{eq:lin13qcd}) yield
\begin{eqnarray}\label{eq:lin21qcd}
\nonumber
\Box'(\bar{\xi}\psi)+a(\bar{\xi}\psi)+b(\bar{\eta}\psi)=0,
\\
\nonumber
\Box'(\bar{\eta}\psi)+a'(\bar{\xi}\psi)+b'(\bar{\eta}\psi)=\Box'(\bar{\eta}\psi)+a'(\bar{\xi}\psi)-a(\bar{\eta}\psi)=0,
\end{eqnarray}
where
\begin{equation}\label{eq:lin22plqcd}
a=a^a T^a,b=b^a T^a,a'=a^a T^a,b'=b'^a T^a,
\end{equation} so
\begin{equation}\label{eq:lin22qcd}
\bar{\eta}\psi=-b^{-1}(\Box'(\bar{\xi}\psi)+a(\bar{\xi}\psi))
\end{equation}
and
\begin{equation}\label{eq:lin23qcd}
(\Box'-a)(-b^{-1})(\Box'+a)(\bar{\xi}\psi)+a'(\bar{\xi}\psi)=0
\end{equation}
or
\begin{equation}\label{eq:lin24qcd}
((\Box'-a)b^{-1}(\Box'+a)-a')(\bar{\xi}\psi)=0.
\end{equation}
Substituting the expressions for $a^a$, $b^a$, $a'^a$ from equation (\ref{eq:lin17qcd}) and using equations (\ref{eq:lin18}) and (\ref{eq:lin24qcd}), we finally obtain:
\begin{equation}\label{eq:lin25qcd}
(((\bar{\xi}\eta^c)\Box'-\bar{\xi}F\eta^c)(\bar{\xi}F\xi^c)^{-1}((\bar{\xi}\eta^c)\Box'
+\bar{\xi}F\eta^c)+\bar{\eta}F\eta^c)(\bar{\xi}\psi)=0.
\end{equation}

This equation for one ``spinor component" $\bar{\xi}\psi$ is generally equivalent to the Dirac equation in the Yang-Mills field (\ref{eq:pr25qcd}) (if $\bar{\xi}F\xi^c$ is not identically singular, which is generally the case): on the one hand, it was derived from the Dirac equation, on the other hand, the field $\psi$ can be restored if its ``spinor component" $\bar{\xi}\psi$ is known and the Dirac equation (\ref{eq:pr25qcd}) can be derived from equation (\ref{eq:pr25qcd}). This can be demonstrated along the same lines as in ~\cite{Akhm2015,Akhmspr}.

The only ``spinor component" in equation (\ref{eq:lin25qcd}) has $n$ ``color" components, however, using a gauge transformation, all of them but one can be made zero and the remaining component  can be made real, as the action of the SU($n$) group on the unit sphere in $\mathbb{C}^n$ is transitive for $n\geq 2$ ~\cite{Knapp}.

\maketitle

\section{\label{sec:level1c}Conclusion}

We considered the Dirac equation in the SU($n$) Yang-Mills field and showed that $3n$ out of $4n$ components of the spinor field can be algebraically eliminated from the equation in a general case, and then all the remaining components but one can be made zero using a gauge transformation, and the last component can be made real. Conceptually, this can be a radical simplification of the Yang-Mills theory.

The resulting equation for the real component describing the spinor field is overdetermined in a general case, so one can hope to eliminate the spinor field altogether, getting equations for the Yang-Mills field only, as such elimination was shown to be possible for scalar electrodynamics ~\cite{Akhmeteli-IJQI,Akhmeteli-EPJC} and spinor electrodynamics ~\cite{Akhmeteli-EPJC}. It would also be interesting to derive the Lagrangian for the Dirac-Yang-Mills system containing only one real component of the spinor field, as such a Lagrangian was derived for spinor electrodynamics (Dirac-Maxwell electrodynamics)~\cite{Akhmetelilagr}.

\section*{Acknowledgments}

The author is grateful to V. G. Bagrov, A. V. Gavrilin, A. Yu. Kamenshchik, T. G. Khunjua, nightlight, and I. V. Ternovskiy for their interest in this work and valuable remarks.

\end{document}